\documentstyle[12pt]{article}

\newcommand\ba{\begin{array}}
\newcommand\ea{\end{array}}
\newcommand\ben{\begin{equation}}
\newcommand\een{\end{equation}}
\newcommand\bea{\begin{eqnarray}}
\newcommand\eea{\end{eqnarray}}

\renewcommand{\theequation}{\arabic{section}.\arabic{equation}}

\def\vev#1{{\langle #1 \rangle}}

\def\math{\mathsurround 0pt}
\def\oversim#1#2{\lower.5pt\vbox{\baselineskip0pt \lineskip-.5pt        
\ialign{$\math#1\hfil##\hfil$\crcr#2\crcr{\scriptstyle\sim}\crcr}}}

\def\lap{\mathrel{\mathpalette\oversim {\scriptstyle <}}}

\def\({\left(} \def\){\right)}
\def\[{\left[} \def\]{\right]}
\def\pa{\partial}
\def\half{{\mathchoice{{\textstyle{1\over 2}}}{1\over 2}{1\over 2}{1 
\over 2}}}

\def\al{\alpha}

\def\ga{\gamma}
\def\de{\delta}
\def\ep{\epsilon}
\def\ze{\zeta}
\def\et{\eta}

\def\ka{\kappa}
\def\la{\lambda}

\def\si{\sigma}
\def\ta{\tau}

\def\Ga{\Gamma}
\def\De{\Delta}

\def\Om{\Omega}

\def\vb{\bar v}
\def\bdi#1{\mbox{\boldmath $#1$}}
\def\bx{{\bdi x}}
\def\by{{\bdi y}}
\def\bX{{\bdi X}}
\def\br{{\bdi r}}
\def\bk{{\bdi k}}
\def\bu{{\bdi u}}  
\def\Xd{\dot X}
\def\Xp{\acute X}
\def\bXd{\dot\bX}
\def\bXp{\acute\bX}

\def\del{\mbox{\boldmath $\nabla$}} 
\def\G{{  G}} 

\def\lah{\la_{\rm h}}
\def\kh{k_{\rm h}}
\def\rs{r_{\rm s}}
\def\etd{\eta_{\rm dec}}
\def\tar{\tau_{\rm r}}

\def\tb{\bar{t}}
\def\des{\De_{\rm sky}}

\def\refjl#1#2#3#4
   {\hangindent=16pt
    \hangafter=1
    \rm #1,
   {\frenchspacing\sl #2
    \bf #3},
    #4\par}

\def\refbk#1#2#3
   {\hangindent=16pt
    \hangafter=1
    \rm #1
   {\frenchspacing\sl #2\/}
    #3\par}

\def\APJ{Ap. J.}
\def\JP{J. Phys.}
\def\MNRAS{MNRAS}  
\def\NP{Nucl. Phys.}
\def\PL{Phys. Lett.}
\def\PR{Phys. Rev.}
\def\PRL{Phys. Rev. Lett.}

\def\cdef#1#2{\edef\doit{\global\def\csname #1\endcsname{#2}}
      \doit}
\def\ct#1{\csname #1\endcsname}

\cdef{AlbTur89}{Albrecht \& Turok 1989}
\cdef{AllShe90a}{Allen  \& Shellard  1990a}
\cdef{AllShe90b}{Allen  \& Shellard  1990b}
\cdef{BarVil89}{Barriola  \& Vilenkin  1989}
\cdef{BenBou89}{Bennett  \& Bouchet  1989}
\cdef{BenBou90}{Bennett  \& Bouchet  1990}
\cdef{BenBouSte93}{Bennett, Bouchet \& Stebbins 1993}
\cdef{BenRhi90}{Bennett  \& Rhie  1990}
\cdef{BenRhi93}{Bennett  \& Rhie  1993}
\cdef{BenSteBou92}{Bennett, Stebbins  \& Bouchet  1992}
\cdef{BenSteBou93}{Bennett, Stebbins  \& Bouchet  1993}
\cdef{BonEfs84}{Bond \& Efstathiou 1984}
\cdef{BonEfs87}{Bond \& Efstathiou 1987}
\cdef{BouBenSte88}{Bouchet, Bennett  \& Stebbins  1988}
\cdef{BraFelMuk92}{Brandenberger, Feldman \& Mukhanov 1992}
\cdef{BraTur86}{Brandenberger \& Turok 1986}
\cdef{CopKibAus92}{Copeland, Kibble \& Austin 1992}
\cdef{Emb92a}{Embacher 1992a}
\cdef{Emb92b}{Embacher 1992b}
\cdef{Fom+93}{Fomalont et al.~1993}
\cdef{Gai+92}{Gaier  et al.~1992}
\cdef{Gan+93}{Ganga, Cheng, Meyer \& Page 1993}
\cdef{Got+90}{Gott et al.~1990}
\cdef{Gun+93}{Gundersen et al.~1993}
\cdef{Kib76}{Kibble 1976}
\cdef{KibCop90}{Kibble  \& Copeland  1990}
\cdef{Mei+93}{Meinhold  et al.~1993}
\cdef{MeyChePag91}{Meyer, Cheng \& Page 1991}
\cdef{MoePerBra94}{Moessner, Perivolaropoulos \& Brandenberger 1994} 
\cdef{MyeReaLaw93}{Myers, Readhead\& Lawrence  1993} 
\cdef{Pee82}{Peebles 1982}
\cdef{PenSpeTur93}{Pen, Spergel \& Turok 1993}
\cdef{Per93}{Perivolaropoulos 1993}
\cdef{Rea+89}{Readhead  et al.~1989}
\cdef{Sch+93}{Schuster et al.~1993}
\cdef{Smo+92}{Smoot  et al.~1992}
\cdef{Ste88}{Stebbins  1988}
\cdef{TraBraTur86}{Traschen, Brandenberger \& Turok 1986}
\cdef{Tur89}{Turok  1989}
\cdef{Vil80}{Vilenkin  1980}
\cdef{Vil85}{Vilenkin  1985}
\cdef{Zel80}{Zel'dovich  1980}

\begin{document}

\begin{titlepage}
\null\vspace{-62pt}
\begin{flushright}
DAMTP--93--17\\
{\tt astro-ph/9307040}\\ 
December 1993 (2nd Revision)
\end{flushright}
\vspace{0.5in}
\baselineskip=36pt
\begin{center}
{{\LARGE \bf Small scale microwave background 
fluctuations from cosmic strings}}\\
\baselineskip=18pt
\vspace{1in}
{\large Mark Hindmarsh\footnote[1]{Present address: 
School of Mathematical and Physical Sciences, 
University of Sussex, Brighton BN1 9QH, U.K.}\\
}
\vspace{0.5in}
{{\it  Department of Applied Mathematics and Theoretical 
Physics,\\
University of Cambridge, Cambridge CB3 9EW, U.K.}} \\

\end{center}
 
\vfill
 
\begin{abstract}

The Cosmic Microwave Background (CMB) fluctuations at very small angular scales 
(less than $10'$) induced by matter sources are computed in a simplified way. 
The result corrects a previous formula appearing in the literature. 
The small scale power spectrum from cosmic strings is then calculated by a
new analytic method. The result compares extremely well with the 
spectrum computed by numerical techniques (when the old, incorrect, formula 
is used).  
The upper bound on the string parameters derived from OVRO data is re-examined, 
taking into account the non-Gaussian nature of stringy perturbations on small 
scales. 
Assuming a conventional ionization history,  the bound is  
$\gamma G\mu < 11\times 10^{-6}$, where $\gamma^2$ is the number of horizon 
lengths of string per horizon volume. Current simulations give 
$\gamma^2 = 31\pm 7$.

\smallskip
\noindent
{\it Subject headings:\/} cosmology: cosmic microwave background --- 
cosmic strings 

\end{abstract}
\end{titlepage}

\baselineskip=24pt

\section{Introduction}
\setcounter{equation}{0}
It is now widely accepted that cosmic microwave background (CMB) fluctuations
test the physics of the very early universe.  With the recent flurry of 
experiments  on various angular scales (\ct{MeyChePag91}; \ct{Smo+92}; 
\ct{Gai+92}; \ct{Mei+93}; 
\ct{Gan+93}; \ct{Sch+93}; \ct{Gun+93}), we are becoming more able to
eliminate theories of the origin of these fluctuations.  There are two main
contenders in the competition to explain the origin of these fluctuations:
quantum fluctuations in the metric during inflation (see for example 
\ct{BraFelMuk92}), and various kinds of semiclassical dynamics after phase 
transitions in field theories (\ct{Kib76}; \ct{Zel80}; \ct{Vil80}; 
\ct{Tur89}; \ct{BarVil89}; 
\ct{BenRhi90}).  Both are able to generate a more or less
Harrison-Zel'dovich  spectrum of density fluctuations, and both produce CMB
fluctuations which are consistent with the COBE observations (\ct{Smo+92}), as 
far as
the theoretical uncertainties allow (\ct{Pee82}; \ct{BonEfs87}; 
\ct{BenSteBou92}; 
\ct{BenRhi93}; \ct{PenSpeTur93}).   The physics is quite different in each
case, and each suffers in different ways from theorists' prejudices. 
However, it is to be hoped that future measurements of the spectrum of CMB
fluctuations on a wide variety of angular scales will be able to distinguish
experimentally between theories. 

This paper is concerned with calculating the very small angular scale
fluctuations from cosmic strings (\ct{BouBenSte88}).  
Strings are perhaps the most venerable
of the theories based on the dynamics of field theories during and after
phase transitions.  Despite predating the inflationary scenario, the theory
has suffered from analytic intractability and is consequently comparatively
undeveloped.  Early work on string seeded galaxy perturbations 
(\ct{BraTur86}; \ct{TraBraTur86}) was based on the first 
numerical simulations (\ct{AlbTur89}) whose
detailed results have not all been confirmed by the two subsequent groups to 
work
on the subject (\ct{BenBou89}; \ct{BenBou90}; \ct{AllShe90a}; \ct{AllShe90b}).  
What is lacking is a good analytic understanding of the
evolution of the string network, although progress is being made in this
regard (\ct{KibCop90}; \ct{CopKibAus92}; \ct{Emb92a}; \ct{Emb92b}).  
It is the intention of this work to try and create an analytic
approach to the calculation of CMB fluctuations from strings.  Although the
results presented apply only to fluctuations on very small angular scales
(less than about 10$'$), 
these are precisely the scales on which strings make their most
distinctive contribution.

The philosophy behind the current approach is quite simple.  It is that
statistical measures of the string network itself can be used to
calculate the statistics of the CMB fluctuations.  One can in fact guess the
form of simple string  correlation functions on general grounds, backed up by
some intuition gained from numerical simulations.   In this paper, 
Stebbins' (1988) formula for small angle CMB anisotropies is rederived 
in a more direct way, with an error in his and previous versions of this work 
corrected (Stebbins 1993). Then, the 
two-point
string correlators are used to calculate the very high frequency end of the
power spectrum of CMB fluctuations.  The result is compared to a
numerical computation of the spectrum by  Bouchet, Bennett and Stebbins
(1988), using the old anisotropy formula.  
In view of the approximations made, it is gratifying that the
current approach reproduces both the shape and the amplitude 
of the spectrum very well.  Limits on the string linear mass
density are then re-examined.  
The results can only be trusted on very small angular scales, 
for which the best experimental limits currently
available are derived from recent VLA observations (\ct{Fom+93}) and from  OVRO 
(\ct{Rea+89}; \ct{MyeReaLaw93}).  
The best experimental geometry for finding string is the recent RING 
experiment at OVRO (\ct{MyeReaLaw93}),  which consists of 96 overlapping 
double difference  
fields in a circle of radius $\sim 2^\circ$ around the North Celestial
Pole. However, sky coverage has been increased at the expense of sensitivity, 
so the best limits on the r.m.s.~fluctuations still come from the NCP 
observations (\ct{Rea+89}). Other experiments at larger scales have
been used in the past to constrain the string scenario, although in view of
the current uncertainty surrounding the predictions of strings these should
not be regarded as reliable. 

The plan of the paper is as follows.  Firstly, I present the rederivation of 
Stebbins' (1988) formula, for small angle microwave anistropies.  
I then use this formula to derive an
expression for the CMB power spectrum, subject to some plausible
assumptions.  These assumptions are  justified by comparison to the numerical
work of  Bouchet, Bennett and Stebbins (1988),  henceforth referred to as BBS.
I then use the derived correlation function to obtain an upper limit on a 
certain combination of string parameters. To do this a revised 
Bayesian analysis is performed on OVRO data, which models 
the non-Gaussian statistics of strings.  
The bound, presented in (\ref{eLim}), is expressed as one on $\gamma G\mu$, 
where $\gamma^2$ is the number of horizon 
lengths of string per horizon volume, for $\gamma$ is not well determined as 
yet.  Furthermore, this combination appears in all calculations of 
string-induced perturbations, and so bounding it is more useful than simply 
bounding the string tension.

\section{CMB anisotropies from moving strings}
\setcounter{equation}{0}
\def\De{\delta}
In this section I derive the formula for the temperature
pattern induced in the CMB by a string moving in front of it.  
A Minkowksi
space background is used, which limits the applicability to fluctuations
inside the horizon at decoupling.
Stebbins (1993) has corrected his original formula (\ct{Ste88}), which 
was also incorrectly derived in earlier versions of this work. We are now, 
happily, in agreement.

Imagine a large box of length $b$ and cross-sectional area $A$.  In this box is
some string with spacetime coordinates $X^\mu(\si,\ta)$, where $\si$ and $\ta$
are respectively spacelike and timelike worldsheet coordinates.  Consider the
gravitational effect of the string on a set of photons passing through this box
with momentum $p_\mu = (E;0,0,E)$.  Their unperturbed geodesics are
\ben
Z^\mu = x^\mu + \la p^\mu.
\een
We would like to know what their energies are at $x^\mu=(t_0;\bx,z_0)$ as a
result of the string's gravitational field.  
To first order in linear theory we have
\ben
\De p_\mu(\bx) = -\half\int_{\la_1}^{\la_0} 
d\la \, h_{\nu\rho,\mu}(Z(\la))p^\nu p^\rho,
\een
where $h_{\mu\nu}$ is the perturbation around the Minkowski metric 
$\eta_{\mu\nu} = {\rm diag}\,(1,-1,-1,-1)$.   

To go further it is very convenient to choose the harmonic gauge, or
\ben
{h^{\mu\nu}}_{,\nu} - \half h^{,\mu} = 0,
\een
where $h=h^\nu_{\;\nu}$.  In this gauge the first order field equations become
\ben
\pa^2 h_{\mu\nu} = 16\pi G(T_{\mu\nu} - \half \eta_{\mu\nu}T),
\een
where $T_{\mu\nu}$ is the stress tensor.  The string has its own 
gauge freedom, which we restrict by choosing the conformal gauge, or
\ben
\Xd \cdot \Xp = 0, \qquad \Xd^2 + \Xp^2 = 0,
\label{eCon}
\een
where the dot and the prime  denote differentiation with respect to $\tau$ and 
$\si$ respectively.  In this gauge the stress tensor is given by
\ben
T^{\mu\nu} = \mu \int d\tau d\si(\Xd^\mu\Xd^\nu - \Xp^\mu\Xp^\nu)\de^{(4)}
\(x-X(\si,\ta)\).
\een
One way to proceed from here is to compute $h_{\mu\nu}$ from $T_{\mu\nu}$ 
with a retarded Green's function, differentiate to get $h_{\nu\rho,\mu}$, 
and finally integrate with respect to the photon affine parameter 
(\ct{Ste88}).  
However, 
a great simplification is afforded by instead calculating $\del^2 \De p_\mu$, 
where $\del$ is the partial derivative with respect to the transverse 
coordinates $\bx$.  This is done in a roundabout fashion, using the fact 
that $(\pa_t -\pa_z)f(Z) = E^{-1}df/d\la$:
\bea
\del^2\De p_\mu &=& (\pa_t^2 - \pa_z^2  - \pa^2)\De p_\mu \nonumber\\
                &=&  - \half\int_{\la_1}^{\la_0} d\la \,
                [(\pa_t+\pa_z)(\pa_t-\pa_z)-\pa^2](h_{\nu\rho,\mu}
                    p^\nu p^\rho)\nonumber\\
                &=& \left[\frac{1}{2E}(\pa_t+\pa_z)h_{\nu\rho,\mu}p^\nu 
                    p^\rho\right]_{\la_1}^{\la_0} - 8\pi G\pa_\mu
                    \int_{\la_1}^{\la_0} d \la \, T_{\nu\rho}p^\nu p^\rho.
\eea
For temperature fluctuations we are interested in the variation on the 
photon energy $\de p_0 \equiv \de E$.  For this component we may use a cunning 
identity due to Stebbins (1993), whose proof is reproduced in the Appendix.
Writing $\hat p^\mu = p^\mu/E$, it is
\ben
\hat p^\nu T_{\nu\rho,0} = - \nabla_\perp^iT_{i\rho} - 
\frac{1}{E}\frac{d}{d\la}\hat 
p^iT_{i\rho}.
\label{eCunId}
\een
Here, $\nabla_\perp^i = (\de^{ij} - \hat p^i\hat p^j)\pa_j $ is the 
transverse derivative.  Thus we have
\bea
\del^2 \frac{\de E}{E} &=& 8\pi G \int_{\la_1}^{\la_0}Ed\la 
\nabla_\perp^i T_{i\rho}\hat p^\rho \nonumber\\
&+& \half\left[(\pa_t +\pa_z)\pa_t\hat h - 16\pi G 
\hat p^i \hat p^\rho T_{i \rho}\right]_{\la_1}^{\la_0}.
\label{eFormula}
\eea
The terms in the square brackets are fluctuations on the bounding surfaces 
of our imaginary box.  Those at the observing time are negligible, since 
we are far from any sources, but if the initial surface is the decoupling 
time, there will be important fluctuations present.  Our justification 
for dropping these is the finite thickness of the last scattering surface, 
which acts to smear out fluctuations on scales greater than about $10'$, 
assuming that $z_{\rm dec} \simeq 1000$.

We now evaluate (\ref{eFormula}) for string sources.  
It is customary in the string literature to choose the temporal 
gauge $X^0 = \tau$; that is, identify the worldsheet time  with the background 
time coordinate.  In that case we find  
(using the fact that $\De T/T = \De E/E$)
\ben
\del^2 \frac{\De T}{T} = -8\pi G\mu \int d\si \(  \bXd - 
{(\Xp \cdot \hat p) \over (\Xd \cdot \hat p)} \bXp \)
\cdot \del \de^{(2)}(\bx - \bX),
\label{eTemGau}
\een
where worldsheet variables are evaluated at 
$t_{\rm r}$, given by $x^+ = X^+(\si,t_{\rm r})$, or $t_{\rm r} = t+z - 
X^3(\si,t_{\rm r})$. Once again, the reader is cautioned that this  formula,
again due to Stebbins (1993), 
corrects Stebbins' (1988) result, and earlier versions of this work. 

The expression is simplified if instead we use 
the light cone gauge  
\ben
X^+(\si,\ta) = \ta.
\label{eLCG} 
\een
The time parameter $\ta$ then labels the intersections with a set of null 
hyperplanes with the worldsheet, which is in hindsight an intuitive thing 
to do, since our photon geodesics $Z^\mu = x^\mu + \la p^\mu$ form just 
such a set.  
Then we find 
\ben
\del^2 \frac{\De T}{T} = -8\pi G\mu \int d\si \Xd\cdot \del  \de^{(2)}(\bx -
\bX), 
\label{eTemFlu}
\een
where quantities are now evaluated at $\ta=x^+$.

These is a nice result: it says that in Minkowski space the temperature 
distortions caused by moving strings depend only on the apparent 
positions of the strings and their light cone gauge transverse peculiar 
velocities, and not on the entire history of the worldsheet.   
In using this gauge we will need to exercise 
care in taking results from numerical simulations, where correlation 
functions are always measured in the temporal gauge.

\section{Power spectrum}
\setcounter{equation}{0}
In this section an expression is derived for the power spectrum of
temperature fluctuations in terms of the two point correlators of the 
transverse coordinates of the string, $X^A(\si)$.  The basic two point
functions are
\ben
\vev{\Xd^A(\si)\Xd^B(\si')},\quad \vev{\Xd^A(\si)\Xp^B(\si')},
\quad  \vev{\Xp^A(\si)\Xp^B(\si')}.
\een
The angle brackets denote an average over an ensemble of strings in our
imaginary box.  The starting assumption is that the string ensemble is a
Gaussian process: that is, all correlators can be calculated  in terms of
the two point functions.  This has not been tested directly, but the results
can be regarded as justifying the means.

We now make some more justifiable assumptions about the ensemble:
(\romannumeral 1) rotation, reflection and translation invariance of the
transverse coordinates;   and 
(\romannumeral 2) worldsheet reflection and
translation invariance.  
Then we can reduce the number of correlation
functions to four: 
\bea
\vev{\Xd^A(\si)\Xd^B(\si')} &=& \half\de^{AB}V(\si-\si'), \\
\vev{\Xd^A(\si)\Xp^B(\si')} &=& \half\de^{AB}M_1(\si-\si') + \half \ep^{AB}
M_2(\si-\si'), \\
\vev{\Xp^A(\si)\Xp^B(\si')} &=& \half \de^{AB} T(\si-\si'). 
\eea
For later convenience two other correlators will be defined: 
\bea
\Ga(\si-\si') &=& \int_{\si'}^\si d\si_1\int_{\si'}^\si d\si_2 T(\si_1-\si_2) 
\equiv \vev{(X^A(\si)-X^A(\si'))^2}, \\
\Pi(\si-\si') &=& \int_{\si'}^\si d\si_1M_1(\si_1-\si) 
\equiv \vev{(X^A(\si)-X^A(\si'))\Xd^A(\si)}.
\eea
$V$ and $T$ are symmetric in their argument, while $M_1$ and $M_2$ are
antisymmetric.  It turns out that we will not need the mixed correlator
$M_2$.

The two point correlation function of the temperature fluctuations is
defined to be
\ben
C(\br) = \vev{\De T(\bx)\De T(\bx+\br)}/T^2
\een
where again the angle brackets denote an average over the string ensemble. In
terms of the two dimensional Fourier transform
\ben
\De_\bk = \int d^2x \frac{\De T}{T}(\bx)e^{\textstyle i\bk\cdot\bx},
\een
we have
\ben
C(\br) = {1\over A} \int {d^2 k\over (2\pi)^2} |\De_\bk|^2
e^{\textstyle -i\bk\cdot\br}.
\een
The power spectrum is the ensemble average of $|\De_\bk|^2$.  The Fourier
transform of the temperature fluctuation formula (\ref{eTemFlu}) is
\ben
-k^2\De_\bk = i8\pi G\mu k^A   \int d\si \Xd^A(\si)
e^{\textstyle i\bk\cdot\bX(\si)}.
\label{eFT}
\een
Thus, normalising for the moment  to unit box area $A$,
\ben
P(k) = (8\pi G\mu)^2 {k^Ak^B\over k^4} \int d\si d\si' 
\left\langle
\Xd^A(\si)\Xd^B(\si')e^{\textstyle i\bk\cdot(\bX(\si)-\bX(\si'))}\right\rangle.
\een
With our assumptions about the string correlation functions the ensemble
average can be reduced to
\ben
P(k) =  \half(8\pi G\mu)^2 {1\over k^2}\int d\si d\si'\left( V(\si-\si') +
\half k^2 \Pi^2(\si-\si') \right) e^{-k^2\Ga(\si-\si')/4}.
\label{ePowSpe}
\een
At this point one should check the Gaussian approximation by measuring $V$,
$M_1$ and $T$ in one's favourite string simulation, computing $P(k)$ from
(\ref{ePowSpe}), and then comparing with $|\De_\bk|^2$ computed by a direct
Fourier transform of the right hand side of (\ref{eFT}) averaged over a set
of string configurations $\{\dot \bX(\si),\bX(\si)\}$.  The latter procedure
was adopted by BBS using a string simulation with the code of Bennett
and Bouchet (\ct{BenBou89}; \ct{BenBou90}).  
Unfortunately, there is very little correlation function data
available, so we will have to content ourselves with making some educated
guesses for the forms of $V$, $M_1$ and $T$, based on visual inspection of
the simulations.  

First, however, we must make the connection between the Minkowski space
formalism and the real behaviour of photon geodesics in an expanding
universe.  Suppose the universe is flat with background metric $g_{\mu\nu}
= a^2(\eta)\et_{\mu\nu}$, where $\eta$ is the conformal time. Our calculations
are valid for comoving momenta and comoving coordinates, and scaled metric
$h_{\mu\nu} = \de g_{\mu\nu}/a^2$, provided we can neglect terms of order
$h_{\mu\nu} \dot{a}/a$ in comparison to $\dot{h}_{\mu\nu}$. This means that
our approximation can only be trusted for perturbations on scales inside the
causal horizon at the time of decoupling of matter and radiation, $\et_{\rm
dec}$.  Bearing that in mind, we would like to infer things about our light
cone gauge correlation functions in a { comoving} box of area $A$ and length
$\eta_0 - \eta_{\rm dec}$.   

Let us outline what is believed about the {\it three\/} dimensional two point
correlators, 
\bea
\vev{\Xp^i(\si)\Xp^i(\si')} &=& \half \G^+(\si-\si'), \\
\vev{\Xd^i(\si)\Xd^i(\si')} &=& \half \G^-(\si-\si'), \\
\vev{\Xp^i(\si)\Xd^i(\si')} &=& {\textstyle {1\over 4}} \Om(\si-\si'), 
\eea
where Embacher's (\ct{Emb92a}; \ct{Emb92b}) notation has been echoed.  
The correlators are all
dimensionless functions, and so must have a scale $\xi$ in them if they are to
be non-trivial.  The evidence from the numerical simulations is that  
this correlation length is
proportional to the most important physical scale in the expanding universe,
the horizon size $\eta$.  Let $s = (\si-\si')/\xi(\eta)$. Then  it is found
that 
\ben
\xi^2 \int_0^s ds_1\int_0^sds_2 \G^+(s_1-s_2) \sim \xi^2 s^{2\nu}.
\een
The exponent $\nu$ is a function of scale:
\ben
\nu(s) \to \cases{\half & if $s \gg 1$, \cr
                1     & as $s \to 0$. \cr} 
\een
That is, on large scales, the network behaves as a Brownian random walk, 
while on very small scales it is approximately straight.  
There is numerical evidence for an ``intermediate fractal'' region for $s
\lap 1$ (\ct{BenBou90}; \ct{AllShe90b}), 
where the exponent slowly interpolates between 1 and 0.5.  

The velocity correlation function $\G^-$ starts out at $2\vb^2$, twice the mean
square velocity, and vanishes rapidly for $s\gg 1$ as befits a random walk. 
There is also a constraint 
\ben
\int_0^\infty ds \G^-(s) = 0
\een
which arises because there can be no coherent velocities in the network above
the horizon scale.  Thus $\G^-(s)$ must go negative somewhere.   The form of
the mixed correlator $\Om$ is also subject to the same integral constraint, and
must also vanish at $s=0$ by its antisymmetry.  The  inferred forms of the
correlators are displayed in Figure 1.

The light cone gauge correlators contain string coordinates evaluated at all
times between $\et_{\rm dec}$ and $\et_0$, and thus have contributions from
the equal time temporal gauge correlators at all times in this interval, as
well as from unequal time ones.  However, most of the  string in the box is
near the initial time surface, so we will assume it is the correlation
length of the string network at $\et_{\rm dec}$ that sets the scale for the
light cone gauge correlators. 

To check that this is true, we note that the physical length density of the
string is proprtional to $\xi^{-2}$, a result of having of order one segment 
of length $\xi$ in a volume $\xi^3$.  If we write $\xi = \eta/\gamma$, where 
$\ga$ is the conformal time, then $\ga^2$ is the number of physical horizon 
lengths of string per horizon volume $\eta^3$, and the comoving length of 
string in comoving volume $Ad\eta$ is
\ben
dL_{\rm c} = \left({\ga^2 \over a(\eta) \et_0^2}\right)Ad\eta,
\een
where $\eta_0$ is toay's conformal time.  In the radiation era, $a(\eta) = 
(\eta/\eta_0)^2$, so
\ben
dL_{\rm c} = \ga^2 {d\eta\over \eta^2}.
\een
Thus most of the string is near $\eta_{\rm dec}$, 
and it is justifiable to assume that the light
cone gauge correlators $T$, $V$ and $M_1$ have the same general form as 
$\G^+$, $\G^-$ and $\Om$ respectively, 
with scale $\xi(\et_{\rm dec})/a(\et_{\rm dec})$.  
The exponent $\nu$ is likely to be greater than 0.5, for the apparent position
of the string need not be precisely Brownian:  the nearer sections of string
are straighter.  

We are now in a position to derive the asymptotic behaviour of $P(k)$  as
$(k\xi)$ gets both large and small.  Let us examine the first term in the
power spectrum (\ref{ePowSpe}),
\ben
P_V(k) = (8\pi G\mu)^2 {1\over 4k^2}  \int d\si_+ d\si_- V(\si_-)
e^{-k^2\Ga(\si_-)/4},
\een
where $\si_\pm = \si\pm\si'$.  For $k\xi \gg 1$ we find
\ben
P_V(k) \simeq  (8\pi G\mu)^2 {1\over 2k^2} L \vb^2 \({4\pi\over
\vev{\bXp^2}}\)^{\half}{1\over k},
\een
where $L$ is the total length of string in the box.   For $k\xi \ll 1$ we
{\em define} $\xi$ such that $\Ga(\si) \to \xi|\si|$ at large $\si$, and find 
\ben
P_V(k) \simeq (8\pi G\mu)^2 {1\over 2k^2} (L\xi) \vb^2 u_1k^2\xi^2,
\label{eLowSpe} 
\een
where $u_1$ is a constant related to a moment of $V$:  
\ben 
u_1 = -{1\over 4\vb^2} \int  |s| V(s)ds.  
\een
The contribution to the power spectrum from the mixed correlator $M_1$ is
\ben
P_M(k) = (8\pi G\mu)^2 {1\over 8} \int d\si_+ d\si_- \Pi^2(\si_-)
e^{-k^2\Ga(\si_-)/4}.
\een
$\Pi^2$ goes as $\si_-^4$ at small $\si_-$, so 
\ben
P_M(k) \sim  (8\pi G\mu)^2(L\xi)(k\xi)^{-5}.
\een
For small $k\xi$, we define another constant $i_1 = -\int  |s| \Pi^2(s)ds /
4\vb^2$ to obtain
\ben
P_M(k) =  {\textstyle {1\over 4}}(8\pi G\mu)^2(L\xi) i_1 k^2\xi^2.
\een
Thus the mixed correlator contributes little at the asymptotes of the 
spectrum, and 
the form of $k^2P(k)$ is now clear.  It rises as $k^2$ for $k \ll \xi^{-1}$,
and falls of as $k^{-1}$ for high spatial frequencies.  This general form 
is to be expected: at low resolution, the temperature pattern is 
uncorrelated,  and so the power spectrum vanishes at low spatial frequency, 
while at high resolution we see a collection of randomly oriented edges.

At this point it must be stressed that the function $P(k)$ that we have been
computing in this section does not represent the only contribution to the
fluctuations in the apparent temperature.  The other contributions include
Doppler shifting from scattering off moving electrons at decoupling, and 
Sachs-Wolf contributions from metric fluctuations in the last scattering
surface and from decaying scalar perturbations created by the motion of the
strings as they enter the horizon between decoupling and the present time.
Furthermore,  the calculation is wrong for $k < \et_{\rm dec}^{-1}$, which 
is a consequence of using the Minkowski space
formalism.  

One could imagine fixing up this latter problem by computing the
contributions to the power spectrum $\De P(k,\et)$ from different
times, cutting off the spectrum for $k<\et^{-1}$, and then adding them
up for all times since $\et_{\rm dec}$.  This approach was taken by
Bennett, Stebbins and Bouchet (1992) to compute the fluctuations on
COBE scales.  This effectively assumes that there are no correlations
between the string positions and velocities at one time and another,
which perhaps not very safe in view of the correlations manifest in
their numerical data (see Section 4).  It also neglects the scalar
perturbations induced by the strings as they fall within the horizon.
One should not necessarily regard the fluctuations computed thus as a
lower bound, for the temperature anisotropies caused by the scalar
modes may well be correlated with those from the discontinuities across
the string.  A related approach in position space was adopted by
Perivolaropoulos (1993). The correlation function was taken to be an
incoherent sum of contributions made each time the universe doubles in
size.  The individual contributions were modelled by a cosine (which is
essentially a geometrical factor arising from assuming that the strings
are perfectly straight within a horizon distance), plus a sharp cut-off
at the horizon.  This form for the correlation function is however
rather different from that derived in Section 5.

Doppler scattering occurs within the last scattering surface, and is a result
of velocity perturbations to the electrons.  These cannot be coherent above
the horizon size at decoupling, which corresponds to an angular scale of about
2$^\circ$.  They are also expected to be smoothed off on a scale corresponding
to the thickness of the last scattering surface, which is about $\De z \simeq
150$ for the conventional ionization history, where $z_{\rm dec} \simeq 1100$.
This corresponds to an angular scale of around 10$'$.  Thus we can trust our
power spectrum  only on scales below this, or $k\et_{\rm dec} \gg 0.1 $.

\section{Comparison with numerical simulations}
\setcounter{equation}{0}
The BBS power spectrum was obtained by solving a Fourier space 
version of the temperature fluctuation formula for several null
slices taken from a single run of a matter era string simulation 
(\ct{BenBou90}), in which the scale factor at the end of the slice was twice 
its inital value. However, the incorrect formula was used, namely
\ben
\del^2\frac{\de T}{T} = -8\pi G\mu \int d\si \bu \cdot \del 
\de^{(2)}(\bx - \bX),
\een
with $\bu = (1 - (\Xp \cdot \hat p)^2/(\Xd \cdot \hat p)^2)\bXd$ 
(all quantities being evaluated in the temporal gauge).  For 
the purposes of testing techniques espoused in this paper, the incorrect 
formula must be used to compare the theoretical prediction with the numerical 
result.  BBS
fitted the power spectrum with a function with four parameters: two for the
asymptotic power laws, one for the position of the maximum, and one for the
overall amplitude.  Their fit is expressed in the integrated form
\ben
\int_{2\pi/\la}^\infty {d^2k\over(2\pi)^2} P(k) =
(6G\mu)^2\({(\la/\lah)^{1.7} \over (0.6)^{1.7} + (\la/\lah)^{1.7}}\)^{0.7},
\label{eBBSSpe} 
\een 
where $\lah$ is the comoving horizon size at decoupling.  This
corresponds to low and high $k$ behaviour of $k^{1.7}$ and $k^{-1.2}$
respectively.  The exponents are not significantly different from the predicted
values of 2 and -1.  To illustrate this, the numerical spectrum has been
compared to the function 
\ben
F(k) = k^2 P(k) / (G\mu)^2{ 2\pi} ,
\label{eFit1}
\een
with  
\ben
F(k) = {a_0(k/\kh)^2 \over [a_1^2+(k/\kh)^2]^{3/2}}, 
\label{eFit2}
\een
$\kh$ being $2\pi/\lah$. Trial and error 
gave $a_0 = 65\pm 5$ and $a_1 = 1.8\pm
0.2$. Figure 2 shows $k^{2}P(k)/2\pi$ for the
numerical spectrum, the BBS fit, and the two parameter fit
(\ref{eFit1},\ref{eFit2}). The errors shown are the standard deviations of
twelve null surfaces in the same simulation, taken at four different times
over an expansion factor of $\sqrt{2}$, in three orthogonal directions.  
They are clearly highly correlated, which makes $\chi^2$ fitting 
pointless.

Not only does the theory predict the  form of the spectrum  at the high and
low frequency ends from the little information that we have about correlation
functions, it also predicts the normalisation.  The
theoretical asymptote of $F$ is, when we return the normalizing area 
$A$ to its rightful place,  
\ben
F_{\rm th}(k) =  16\sqrt{\pi}\frac{\vb^2}{\tb}\({L\lah\over A}\) 
\({\kh\over k}\),
\een
where $\tb^2 = {\vev{\bXp^2}}$, 
which we would like to compare to the numerical fit $a_0(\kh/k)$.  
Here we run into a problem engendered by using the light cone gauge, for 
the numerical simulations do not directly give $\tb$, $\vb$ or $L$. 
We have to convert to the temporal gauge, in which
\ben
\vb^2 \to \vev{\bu^2(\tar)}, \qquad \tb \to \surd(\vev{\bXp^2(\tar)}),
\een
The mean square velocity $V^2 = \vev{\Xd^i\Xd^i}$ 
in the matter era in the Bennett and Bouchet simulations is 
$0.37\pm 0.02$ (\ct{BenBou90}).  
Since the temperature fluctuations depend only on the 
transverse components,  $\vev{\bu^2}$ must be strictly less than  
${2\over 3} V^2 $.  
The exact figure cannot be calculated with the information at hand, but 
we can make a crude guess by 
replacing $(\Xp^3)^2$ and $ (\Xd^3)^2 $ by their mean values, and ignoring 
all other correlations.   We also face the problem of evaluating terms like 
$\vev{(\Xd\cdot\hat p)^{-m}}$, with $p = 2,4$.  We cannot consistently use 
a Gaussian approximation here, since these expectation values diverge due 
to the contribution at $\Xd^3 = -1$, where the string has a cusp moving 
towards the observer.  However, we should not expect a Gaussian distribution 
for quantities like $\Xd^3$ which are constrained to be less that or equal to 
1.  Instead, we shall do the simplest thing, which is a series expansion 
in $\vev{(\Xd^3)^2}$, so that
\ben
\vev{(\Xd\cdot\hat p)^{-m}} \simeq 1 + \half m (m+1) \vev{(\Xd^3)^2}.
\een
We can evaluate the required correlators with one more piece of information, 
the temporal gauge constraint $\vev{(\Xd^i)^2} + \vev{(\Xp^i)^2} = 1$. Then
\bea
\vev{(\Xp^3)^2} = \frac{1}{3}(1-V^2), &\qquad&
\vev{(\Xd^3)^2} = \frac{1}{3} V^2, \\
\vev{\bXp^2}    = \frac{2}{3} (1-V^2), &\qquad&
\vev{\bXd^2}    = \frac{2}{3} V^2.
\eea
Thus we find
\ben
\vev{\bu^2} = \frac{2}{3}V^2\(1 - \frac{2}{3}(1-V^4)+\frac{1}{3}
(1-V^2)^2(1+\frac{10}{3}V^2)\).
\een
Putting in the numerical value of $V^2$, we have
\bea
\vb^2 &=& 0.18 \pm 0.01, \\
\tb^2 &=& 0.42 \pm 0.02. 
\eea
The total
projected comoving length of string per unit area is  
\ben
\({L\over A}\) =  \lah^{-1}[1-  (z/z_{\rm dec})^{1/2})] \gamma^2
\een
where $\gamma^2 = {\rho_{\rm s} \lah^2/ \mu z_{\rm dec}}$ 
is the number of horizon lengths of string
per horizon volume (\ct{CopKibAus92}), which is $31\pm 7$ in the matter era 
simulations of Bennett and Bouchet (\ct{BenBou90}). 
The BBS spectra have $z/z_{\rm dec}$ = 1/2, so   
\ben
L\lah/A \simeq  9.1  \pm 2.1.
\een
Thus the asymptotic theoretical spectrum is   
\ben
F_{\rm th}(k) \simeq (72 \pm 17) (\kh/k), 
\label{eThSp}
\een
which is in remarkably good agreement with the numerical spectrum.  The 
assumptions that have been made amount to approximating the source of the small 
scale fluctuations by uncorrelated straight segments of moving string, which 
we see reproduces the spectrum very well.   
The error quoted in (\ref{eThSp}) 
is statistical, and mostly due to fluctuations in the string
density between different simulations.   

Lacking data for $V(s)$, no prediction can be made for the low frequency 
end.  Instead   a hostage to fortune can be created by inferring a value for 
$u_1$ through comparison of the low frequency end of the fitted  spectrum 
(\ref{eFit2}) with its theoretical form (\ref{eLowSpe}).  We find 
\ben
u_1 = \sqrt{4\pi}/\tb(\xi\kh a_1)^3.
\een
The correlation length $\xi$ can be estimated from Fig.~2 of Bennett 
and Bouchet (1989) to be about $0.2\lah$ in the matter era.  Thus 
the prediction for $u_1$ is 
\ben
u_1 \simeq 0.1,
\een
whose confirmation (or otherwise) will have to await more detailed 
measurements of string correlation functions.

\section{Comparison with observation at small angular scales}
\setcounter{equation}{0}
\def\De{\Delta}
The best experiment which picks out the small scales where the theoretical
spectrum can be trusted is OVRO (\ct{Rea+89}; \ct{MyeReaLaw93}), which   
is a double difference experiment, with
FWHM $108''$ and beam throw $7'.15$.  
There are also some recent VLA observations (\ct{Fom+93}), 
which are of two regions
about $7'$ across with up to $10''$ resolution.  However, the  
the VLA does not currently have as good  sensitivity as OVRO. The current
theoretical uncertainties in the spectrum above $10'$ make any bounds derived
from larger scale experiments unreliable, and it is difficult to even quantify
the level of unreliability.  

The correlation function of the temperature fluctuations from strings is given
by  
\ben
C(\br) = (G\mu)^2\int {d^2k\over 2\pi k^{2}}F(k)e^{\textstyle i\bk\cdot\br}
\een
This is observed by apparatus with a finite resolution.  If $B(\bx)$ denotes
the beam response, then the observed temperature distribution is
\ben
T_{\rm obs}(\bx) = T*B = \int d^2yT(\bx+\by)B(\by),
\een
and so the observed correlation function must also be convolved with the beam
response: 
\ben
C_{\rm obs} = C*(B*B).
\een
The OVRO beam response is well approximated by a Gaussian (\ct{Rea+89}),
\ben
B*B = {1\over 4\pi r_0^2} \exp(-r^2/4r_0^2),
\een
where $r_0 = 0'.4247$.  We denote the correlation
function smeared on a scale $r_0$ by $C(r_0,r)$.  The experiment measures the
difference in temperature between a central beam ($T_{\rm M}$) and the average
of two flanking ones ($T_{\rm R1}$ and $T_{\rm R2}$), separated by $\rs =
7'.15$. The signal is therefore a temperature difference
\ben
\De T = T_{\rm M} - \half (T_{\rm R1} + T_{\rm R2}),
\een
and so the the mean square temperature fluctuation measured by this type of
experiment is
\ben
\vev{(\De T/T)^2} = \frac{3}{2}C(r_0,0) - 2C(r_0,\rs) + \frac{1}{2}
C(r_0,2\rs).
\een
There is a scale length $\ze = 1/\kh a_1$ in the power
spectrum,  which at $\sim 0.1 \lah \simeq 10'$ 
is about half the three dimensional 
correlation length of the string network, as defined before 
Eq.~(\ref{eLowSpe}).  
Defining a dimensionless wavenumber $\ka = k \ze$, we
find 
\ben 
C(r_0,r) =   \int {d\ka\over \ka} F(\ka)J_0(\ka r/ \ze)
\exp(-\ka^2r_0^2/ \ze^2), 
\een
where $J_0$ is the zeroth order Bessel function.  
We take the power spectrum to be
\ben
F(k) = {A_0 \over a_1}{(k\ze )^2\over[1 + (k \ze)^2]^{3/2}}.
\label{eAnsatz}
\een
We must now give the spectrum its correct normalization, using (\ref{eFormula}).
Hence we have 
\ben
A_0 = 16\sqrt{\pi}\frac{1}{\tb}\vev{(\bXd - (\Xd\cdot \hat p)/(\Xp\cdot\hat p)
\bXp)^2} [ 1 - (z/z_{\rm dec})^\half ] \gamma^2,
\een
which we can evaluate as before, including all the string between the last 
scattering and the present day.  The quantity in the angle brackets turns 
out to be $0.37 \pm 0.02 $, so we arrive at
\ben
A_0 = (16.2 \pm 1.3)\ga^2.
\een
The form (\ref{eAnsatz}) is very convenient, because an
approximation to the integration, valid when $(r_0/\ze)^2 \ll 1$,
can be found in tables.  Plugging in this function into the expression for
$C(r_0,r)$, we find 
\ben 
C(r_0,r) \simeq C(0,r) + \half C''(0,r) (r_0/\ze)^2,
\een
where
\ben
C(0,r) = {A_0\over a_1} 
\exp(-r/\ze).\label{eCorFun} 
\een
Thus   
\ben
\vev{(\De T/T)^2}(G\mu)^{-2} \simeq  
{3\over 2}{A_0\over a_1}\(1- \frac{4}{3}e^{-\rs/\ze}+
\frac{1}{3}e^{-2\rs/\ze}\) + O(r_0^2/\xi^2).  
\label{eDTSq}
\een
For small $\rs/\ze$, as is relevant in a reionized universe, 
the right hand side is a factor ${  \rs /  \ze}$ down on the total 
mean square fluctuation.  One can interpret this factor as 
the probability that the beam pattern will straddle a string. 

Putting in the experimental value for $r_s$, the theoretical value 
of $A_0$, and the fitted value of $a_1$, we find that the expected 
r.m.s.~temperature fluctuations for OVRO are, with $z_{\rm dec} = 1100$,   
\ben
\vev{(\De T/T)^2} \simeq (6.3\pm 0.5)(\ga G\mu)^2.
\label{eThDTSq}
\een
We recall that the error is a  statistical one corresponding to 
1$\si$ fluctuations in mean square string velocity.  When $z_{\rm dec} 
\ll 1100$, the formula (\ref{eDTSq}) gives
\ben
\vev{(\De T/T)^2} = A_0  (2\pi r_s / \lah) (G\mu)^2 \simeq (7.1\pm 0.6)
(\gamma G\mu)^2 (z_{\rm dec}/1100)^\half.
\een

The OVRO NCP 95\% confidence limit on the residual sky variance is (\ct{Rea+89})  
$\De T_{\rm sky}< 58\;\mu$K, or $\des = \De T_{\rm sky}/T_{\rm sky} 
< 2.1\times 10^{-5}$ with $T_{\rm sky} = 2.73$.  This was 
obtained using a Bayesian analysis with a uniform prior on $\des$, 
assuming that the sky fluctuations were uncorrelated between the fields 
and Guassian.  The first assumption is good for string sources too, 
since the NCP fields are separated by $30'$ and the scale length of 
the correlation function at short distances is $\ze \simeq 10'$. 
However, the non-Gaussian nature of string-induced fluctuations on small scales 
means that we must repeat the analysis, using a stringy probability 
distribution $P_{\rm s} $ for $\De T$.  This is well approximated 
by an exponential function (\ct{Got+90}; \ct{BenBouSte93}).  The theoretical 
analysis of Moessner, Perivolaropoulos and Brandenberger (1994) also seems 
to support this contention.  Thus the probability density for measuring 
$\De$ in an observation with noise $\si$ is
\ben
P(\De,\si) = [{2\sqrt{\pi}\si\des}]^{-1}\int d\De' \exp( - 
\sqrt{2}|\De-\De'|/\des + \De'/2\si^2).
\een
The likelihood function for the seven uncontaminated 
measurements $\{\De_i,\si_i\}$, which 
can be found in Table 4 of Readhead et al (1989), is then 
\ben
L(\{\De_i,\si_i\}|\des) = \prod_i P(\De_i,\si_i)
\een
This function, normalized to a maximum value of 1, is plotted in Figure 3. 
Assuming a uniform prior, as do Readhead et al (1989), 
it gives the Bayesian probability density 
for $\des$.  From it we can infer that
\ben
p(\des < 2.8\times 10^{-5}) = 0.95,
\label{e95Conf}
\een
which is conventionally interpreted as a 95\% confidence limit.  We note that 
this limit is significantly looser than the bound on Gaussian fluctuations. 
This is because $P_{\rm s}$ is strongly peaked near $\De T =0$.

We can now bound $\ga G\mu$: using our 95\% confidence limit (\ref{e95Conf}), 
and the central theoretical prediction (\ref{eThDTSq}), 
we obtain (for a universe without reionization)
\ben
\ga G\mu < 11 \times 10^{-6}.
\label{eLim}
\een
We cannot rigorously bound the string tension, for the probability 
distribution of $G\mu$ depends crucially on how the probability distribution 
of $\ga$, say $P_\ga$, behaves near $\ga=0$. The probabilities for $\De$ and 
$\ga$ are independent, so
\bea
P(G\mu) &=& \int_0^\infty P_{\rm s}(6.3\ga G\mu)P_\ga(\ga)\ga d\ga,\nonumber\\
        &=& {1\over (6.3G\mu)^2}\int_0^\infty P_{\rm s}(\De)P_\ga(\De/6.3G\mu)
        \De d\De.
\eea
If $P_\ga$ does not vanish sufficiently fast as $\ga\to 0$, then $P(G\mu)$ will
not converge swiftly enough to have a variance.  Since we do not know $P_\ga$, 
it is probably not worth doing more than estimating a bound for $G\mu$.
The $2\si$ statistical lower bound on $\gamma$ is 4.1, so  
at a kind of 90\% confidence level, we can say
$
G\mu < 3 \times 10^{-6}.
$
One is entitled to question the accuracy of the estimates of the correlation 
functions. However, the estimation method reproduced the numerical power 
spectrum to within about 10\%, which corresponds to about 5\% in the calculation 
of $\ga G\mu$.  Thus for safety we should perhaps quote 
$\ga G\mu < 12\times 10^{-6}$.  A more precise 
bound will have to await a thorough reanalysis of the BBS simulations 
(\ct{BenBouSte93}).  However, this is still very stringent, and depends rather 
weakly on the decoupling redshift: the bound is raised by a factor 
$(1100/z_{\rm dec})^{1/4}$.

We conclude by noting that the RING experiment in particular  
(\ct{MyeReaLaw93}) deserves close
examination, for there will be several fields containing strings (if the
strings are there of course). This experiment does in fact report a signal,
which if fitted to a Gaussian distribution results in an excess variance of
around $100\mu$K. Although the authors do not entirely exclude contamination
by point sources, it would be interesting to test the hypothesis that this
signal is due to string. \def\De{\Delta}

\section{Conclusions}
\setcounter{equation}{0}
In this paper a better analytic understanding of the CMB fluctuations
produced by strings has been arrived at, at least on small angular scales
(less than $10'$).  There are a number of ingredients in the success of the
approach, which unfortunately make its extension to larger scales
difficult.  Principally, the Minkowski space approximation results in a very
simple formula for the projected temperature pattern, which depends only on
the positions and velocities of the strings on the backward light cone of the
observer.  This can only be justified for fluctuations on angular scales
less than a degree.  An encouraging success is that the Gaussian approximation 
for strings, in which only the two point
string correlators are used, reproduces accurately the numerical spectrum 
computed by BBS.  Using the (corrected) theoretical spectrum, 
a limit (\ref{eLim}) can be derived on $\ga G\mu$ from observational data 
on small angular scales, where $\ga$ is the number of horizon lengths of string 
per horizon volume. Because of the corrected formula, this translates to a more 
stringent limit than that given in Bennett, Bouchet \& Stebbins (1989), even 
when statistical and theoretical uncertainties are properly accounted for.
We also find that the correlation function on very small scales is 
approximately 
$C(r) \simeq C(0)\exp(-r/\ze)$, with $\xi \simeq 10'$  
(see Eq.~\ref{eCorFun}).  This is rather different from the   
Standard Cold Dark Matter form $C(0)/(1+r^2/2\al^2)$, with $\al \simeq 10'$ 
(\ct{BonEfs84}). In experimental papers, a
Gaussian autocorrelation function is often used to derive limits on 
temperature fluctuations, which is a reasonable approximation to the 
SCDM correlation function.  However, when limits on cosmic strings 
are required, an exponential correlation function is to be preferred.  
 
I am extremely grateful to Albert Stebbins for many useful discussions, and 
particularly for communicating the revised formula for temperature 
anisotropies.  I am also grateful 
to him and his collaborators David Bennett and Fran\c{c}ois Bouchet 
for making available their data.
I thank also Paul Shellard for discussing his and Bruce Allen's numerical 
results, and Ron Horgan for a factor $\log_{ e} 10$.  
This work was suported by the SERC.  

\section{Appendix}
\renewcommand{\theequation}{A\arabic{equation}}
\setcounter{equation}{0}
Our purpose is here to establish the identity 
\ben
\hat p^\nu T_{\nu\rho,0}(Z) = - \nabla_\perp^iT_{i\rho}(Z) - 
\frac{1}{E}\frac{d}{d\la}\hat p^iT_{i\rho}(Z),
\een
where $Z^\mu = x^\mu + \la p^\mu$.
used in Section 2 to derive the anisotropy formula.  This we do using 
energy-momentum conservation, for
\bea
\hat p^\nu T_{\nu\rho,0} &=& T_{\rho 0,0} + \hat p^i T_{\rho i,0}, \nonumber\\
                         &=& T_{\rho j,j} + \hat p^i T_{\rho i,0}.
\eea
Now we use the fact that when acting on functions of $Z$,
\ben
\pa_t = \frac{1}{E} \frac{d}{d\la} - \hat p^i\pa_i.
\een
Thus
\bea
\hat p^\nu T_{\nu\rho,0} &=& T_{\rho j,j} + \hat p^i \hat p^j T_{\rho i, j}
 + \frac{1}{E}\frac{d}{d\la}\hat p^iT_{\rho i},\nonumber\\
&=& (\de^{ij} - \hat p^i \hat p^j)\pa_j T_{\rho i, j} - 
\frac{1}{E}\frac{d}{d\la}\hat p^iT_{i\rho}, 
\eea
which establishes the result, for $\de^{ij} - \hat p^i \hat p^j$ projects 
onto the transverse coordinates.

\vfill\eject

\section{References}
{\parindent 0pt
\refjl{Albrecht A. \& Turok N. 1989}{\PR}{D40}{973}
\refjl{Allen B. \& Shellard E.P.S. 1990a}{\PRL}{64}{119}
\refbk{Allen B. \& Shellard E.P.S. 1990b, in}{Formation and Evolution of 
Cosmic Strings}{ed. Hawking S.W., Gibbons G. \& Vachaspati T. (Cambridge: 
Cambridge Univ. Press)}
\refjl{Barriola M. \& Vilenkin A. 1989}{\PRL}{63}{341}
\refjl{Bennett D.P. \& Bouchet F.R. 1989}{\PRL}{63}{2776}
\refjl{Bennett D.P. \& Bouchet F.R. 1990}{\PR}{D41}{2408}
\refbk{Bennett D.P. \& Bouchet F.R. \& Stebbins A. 1993,}
{{\rm private communication}}{}
\refjl{Bennett D.P. \& Rhie S.H. 1990}{\PRL}{65}{1709} 
\refjl{Bennett D.P. \& Rhie S.H. 1993}{\APJ}{406}{L7} 
\refjl{Bennett D.P., Stebbins A. \& Bouchet F.R. 1992}{\APJ}{399}{L5}
\refjl{Bond J.R. \& Efstathiou G. 1984}{\APJ}{285}{L45}
\refjl{Bond J.R. \& Efstathiou G. 1987}{\MNRAS}{226}{655} 
\refjl{Bouchet F.R., Bennett D.P. \& Stebbins A. 1988}{Nature}{335}{410} 
\refjl{Brandenberger R.H., Feldman H.A. \& 
Mukhanov V.F. 1992}{Phys. Rep.}{215}{203}
\refjl{Brandenberger R. \& Turok N. 1986}{\PR}{D33}{2182}
\refjl{Copeland E.J., Kibble T.W.B. \& Austin D. 1992}{\PR}{D45}{R1000}
\refjl{Embacher F. 1992a}{\NP}{B387}{129}
\refjl{Embacher F. 1992b}{\NP}{B387}{163}
\refjl{Fomalont E.B. et al. 1993}{\APJ}{404}{8}
\refjl{Gaier T. et al. 1992}{\APJ}{398}{L1}
\refjl{Ganga K., Cheng E., Meyer S. \& Page L. 1993}{\APJ}{410}{L54}
\refjl{Gott J.R. et al.~1990}{\APJ}{352}{1}
\refjl{Gundersen J.O. et al. 1993}{\APJ}{413}{L1}
\refjl{Kibble T.W.B. 1976}{\JP}{A9}{1387}
\refbk{Kibble T.W.B. \& Copeland E.J. 1990, in}{Nobel Symposium 79: The Birth 
and Early Evolution of our Universe}{ed. Gustafsson B., Nilsson Y.-S. \& 
Skagerstam B.-S. (Singapore: World Scientific)}
\refjl{Meinhold P. et al. 1993}{\APJ}{409}{L1}
\refjl{Meyer S., Cheng E. \& Page L. 1991}{\APJ}{371}{L7}
\refbk{Moessner R., Perivolaropoulos L. \& Brandenberger R. 1994,}{{\rm \APJ, 
to appear}}{}
\refjl{Myers S.T., Readhead A.C.S. \& Lawrence C.R. 1993}{\APJ}{405}{8}
\refjl{Peebles P.J.E. 1982}{\APJ}{263}{L1}
\refbk{Pen U.L., Spergel D.N. \& Turok N. 1993, }{{\rm 
Princeton preprint}}{PUP-TH-1375}
\refjl{Perivolaropulos L. 1993}{\PL}{B298}{305}
\refjl{Readhead A.C.S. et al. 1989}{\APJ}{346}{566}
\refjl{Schuster J. et al. 1993}{\APJ}{412}{L47}
\refjl{Smoot G. et al. 1992}{\APJ}{396}{L1}
\refjl{Stebbins A. 1988}{\APJ}{327}{584}
\refbk{Stebbins A. 1993,}{{\rm private communication}}{}
\refjl{Traschen J., Brandenberger R. \& Turok N. 1986}{\PR}{D34}{919}
\refjl{Turok N. 1989}{\PRL}{63}{2625}
\refjl{Vilenkin A. 1980}{\PRL}{46}{1169; erratum, {\frenchspacing\sl \PRL}, 
{\bf 46}, 1496}
\refjl{Zel'dovich Ya.B. 1980}{\MNRAS}{192}{663}
}

\vfill\eject

\section{Figure captions}

\parindent=0pt 
{\bf Figure 1.} \smallskip
The form of the three dimensional two-point string correlation functions 
$\G^+(\si) = 2\vev{\Xp^i(\si)\Xp^i(0)}$ (Figure 1a); $\G^-(\si) =
2\vev{\Xd^i(\si)\Xd^i(0)}$ (Figure 1b); and $\Om(\si) =
4\vev{\Xp^i(\si)\Xd^i(0)}$ (Figure 1c). 

\bigskip 
{\bf Figure 2.}
\smallskip Numerical and analytic forms of the power spectrum of CMB
fluctuations induced at small angular scales by cosmic strings between 
$z_{\rm dec}$ and $z_{\rm dec}/2$.  The spectrum is defined by $F(k) = 
k^2 \vev{|\de_\bk|}^2/2\pi(G\mu)^2$, and plotted against
wavenumber in units of $k_{\rm h} = 2\pi/\etd$, where $\etd$ is the horizon
size at decoupling.  The fitted functions are given in equations
(\ref{eBBSSpe}) and (\ref{eFit2}).  

\bigskip
{\bf Figure 3.}
\smallskip
The likelihood function for the OVRO NCP measurements (excluding one 
contaminated field), using an exponential model for the probability 
distribution of string-induced sky fluctuations.  This gives 
$p(\De T_{\rm sky} < 77\;\mu K) = 0.95$, which is a bound of 
$2.8 \times 10^{-5}$ on the fractional temperature anisotropy.  This is 
about a factor of $4/3$ greater than the Gaussian limit.

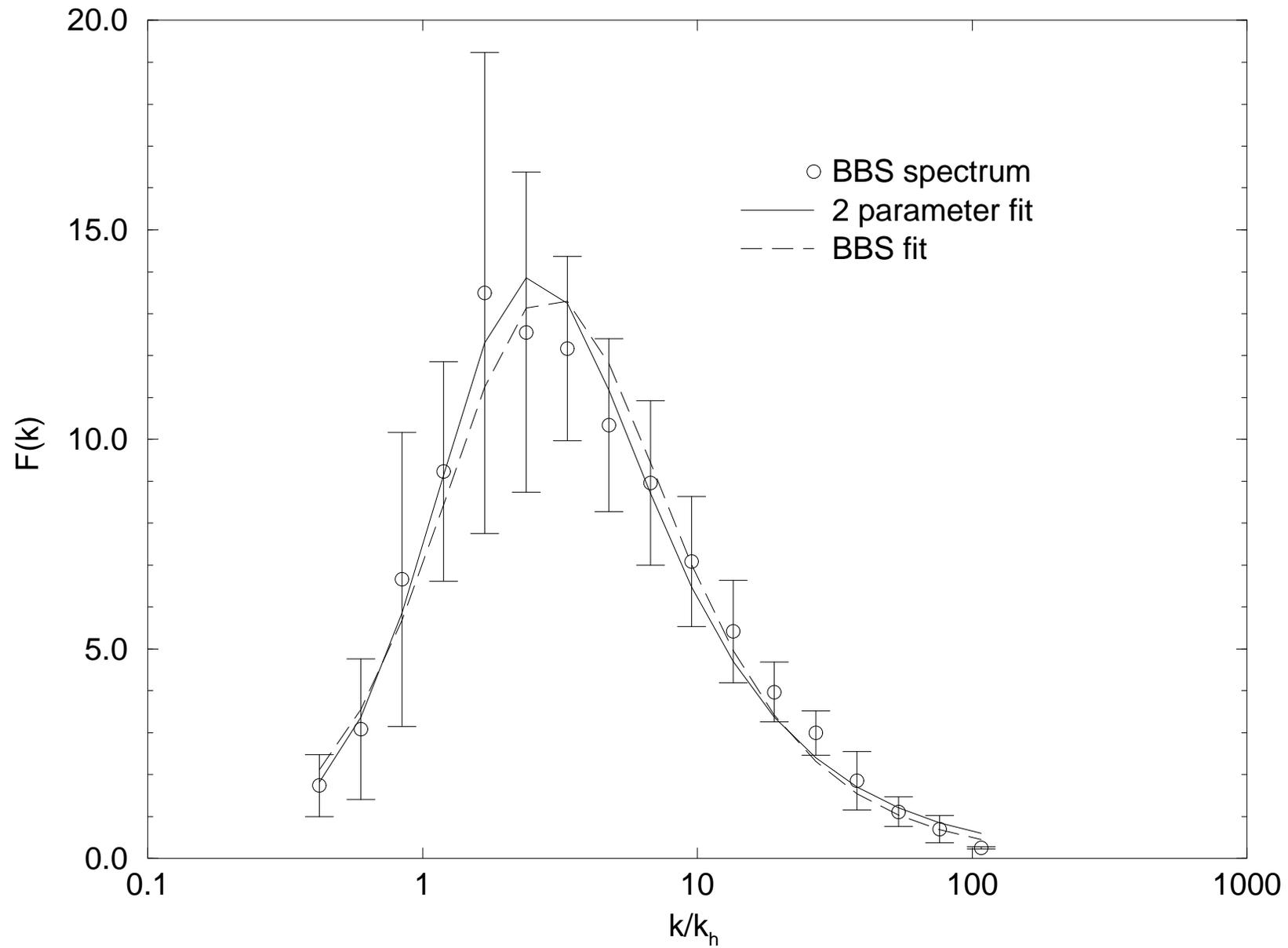
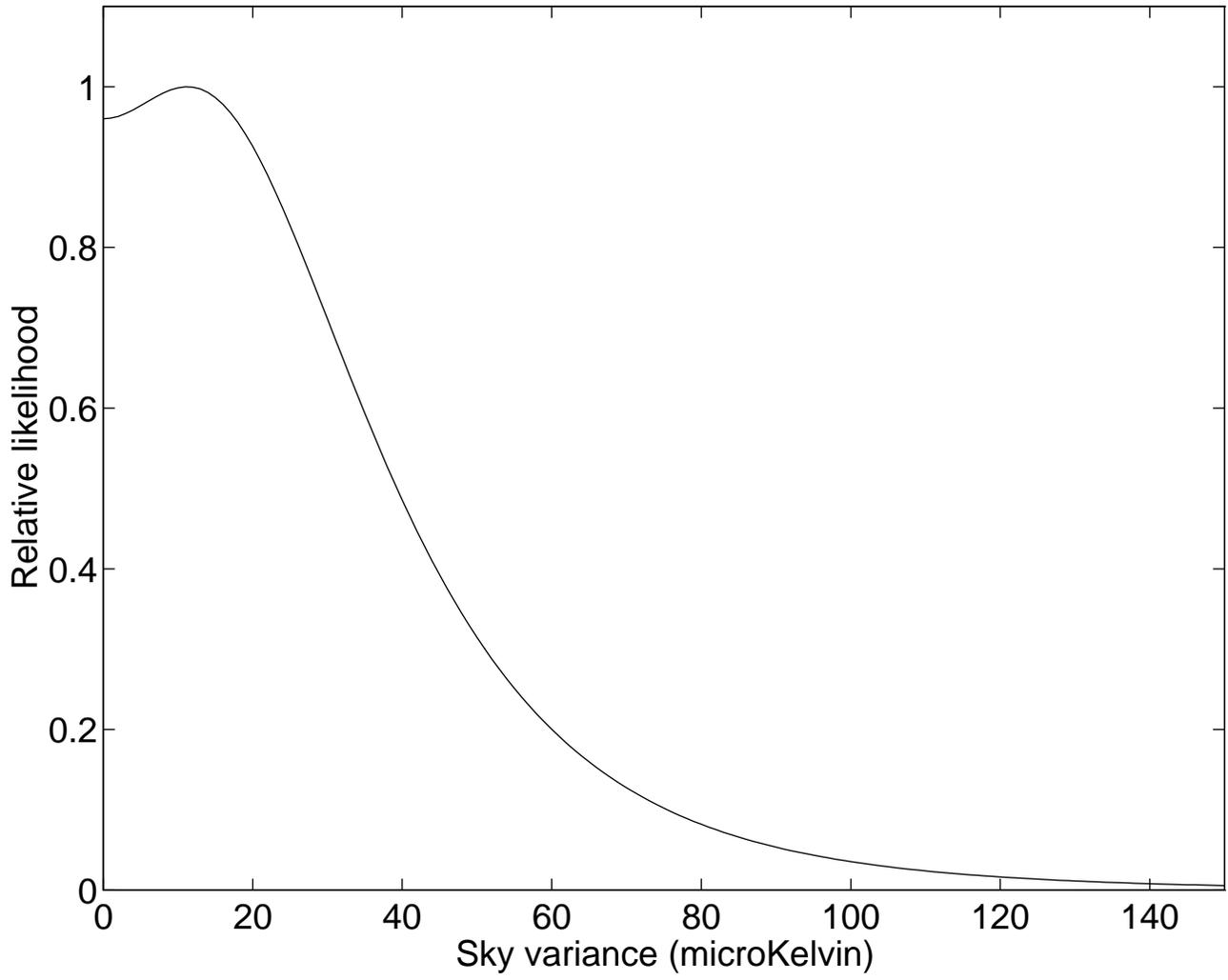
\end{document}